\documentclass[12pt]{article} 
\usepackage[dvipdfmx]{graphicx,xcolor}

\newcommand{\apj}{{\it Astrophys.\ J.}}

\newcommand{\pasj}{{\it Publ.\ Astron.\ Soc.\ Japan}}

\newcommand{\suzaku}{{\it Suzaku}\ }

\newcommand{\chandra}{{\it Chandra}\ }
\newcommand{\xmm}{{\it XMM}\ }

\newcommand{\astroh}{ASTRO-H\ }

\newcommand{\citet}[1]{\cite{#1}}
\newcommand{\citep}[1]{\cite{#1}}

\let\aj=\AJ
\let\aap=\AAP
\let\apj=\APJ

\let\pasj=\PASJ
\let\mnras=\MNRAS

\let\citep=\cite
\let\citet=\cite

\newcommand{\xrism}{{\it XRISM}\ }
\newcommand{\hitomi}{{\it HITOMI}\ }
\newcommand{\kms} {km s$^{-1}$ }

\newcommand{\physrep}{{\it Physics Report}}

\def\vbulk{$v_\mathrm{bulk}$}
\def\vdis{$\sigma_\mathrm{dis}$}

\begin{document}
%
\centerline
    {\large{\bf
{ \\ 
  Gas velocity 
  in the merging cluster Abell 2256
  \footnote{submitted to the XRISM GO1 proposal.
Takayuki TAMURA et al. \today}
}}}  

\bigskip

\section{Merging clusters X-ray spectroscopy}
\subsection{Cluster dynamics and gas velocity}
\label{1-intro:1}
Galaxy clusters 
form hierarchically through collisions and mergers of smaller systems.
Spatial and radial velocity distributions
of member galaxies and X-ray observations of the intracluster medium (ICM)
have revealed that some systems are still forming and un-relaxed.
Sharp X-ray images
revealed shocks and density discontinuities (``cold fronts'') 
suggesting supersonic or transonic gas motions, respectively. 
These heating processes would develop gas turbulence,
cascading down from the driving scales to dissipative ones,
which are determined by unknown viscosity.
These are assumed to accelerate particles, reorder and amplify the magnetic fields,
and generate diffuse radio halos and relics.
To understand these energy flows,
the gas motions in a range of scales should be measured.

\subsubsection{X-ray gas velocity CCD measurements}
The gas bulk motion can be measured most directly using X-ray line emission Doppler shift.
These measurements have been challenging due to the limited energy resolutions of current X-ray instruments.
Typical CCD energy resolution is about 130~eV (FWHM), 
compared with a possible energy shift of $\sim 20$~eV corresponding to a radial velocity of 1000 \kms,
all at the Fe-K line energy here and hereafter.

\cite{2014ApJ...782...38T} found a hint of gas bulk motion (\vbulk ) at $<100$~kpc west of the Perseus cluster center,
where a cold front is seen in X-ray images.
\cite{2016PASJ...68S..19O}
reported 
\suzaku measurements of gas \vbulk\,  
in 8 clusters without significant signs of velocity variations.
We also used \xmm PN spectra
to measure gas \vbulk\, maps
in the Perseus and coma \citep{2020A&A...633A..42S},
A~3667 \citep{2024arXiv240310150O},
and others. 

\subsubsection{
X-ray imaging and \xmm RGS limits}
X-ray brightness (ICM density) fluctuation analyses are used to study gas dynamics
(e.g., Zhuravleva et al. 2014).
Furthermore, the XMM RGS are used to limit gas dynamics
based on line broadening
or resonant scattering.

\subsection{High energy resolution spectroscopy}
           {\bf Hitomi Perseus results
(\cite{2018PASJ...70....9H})
:}
We used the Hitomi $\sim 5$~eV resolution spectra to 
resolve emission lines and 
measure the gas Doppler shifts and width.
Within the central 100~kpc core, 
we found a line of sight (LOS) velocity gradient 
of $100-200$ km s$^{-1}$.
This could be due to the buoyant AGN bubble or central gas sloshing.
A velocity dispersion (\vdis) is also revealed.
These velocities correspond to a galaxy scale potential
and $<10$\% of the ICM thermal pressure.
We also used resolved X-ray lines of various elements and ions
to limit the non-Gaussianity of the line shape, 
test ion-electron thermal equilibrium, 
and resonant scattering.
These observations are limited to the Perseus core. 

From the \hitomi analysis, 
we learned that high resolution spectra
include a wealth of information but are complex.
For example,
the line profiles depend not only on small to large-scale motions
but also on 
the atomic and thermal nature of the ICM projected onto the telescope beam.

X-ray imaging studies are useful but require specific collision configurations to be applied to observations.
In contrast, 
X-ray Doppler shift measurements 
are more direct and commonly applicable to merging systems.
This method is sensitive to the motion parallel to the LOS.
These two measurements are complementary and could provide a direct measurement of three-dimensional motion
if applied to a merging system simultaneously.

\subsection{Target selection and Abell 2256 }
{\bf General motivations:}
In almost all cases we found that X-ray gas redshift is consistent with
those of optical galaxies around the corresponding region.
However, previous measurements are limited mostly to X-ray bright and relaxed regions.
In systems under a violent phase,
we expect to find segregation in the location and velocity
of the gas and galaxies.
Such detachments in the sky position have been observed in some merging systems such as the Bullet cluster.

To understand cluster formation,  which is dominated by non-linear processes, 
systematic measurements in a sample of clusters are required.
In \cite{2014arXiv1412.1176K},
we examined and proposed various types of bright targets
for \astroh studies of clusters.
Five merging clusters were selected, including \xrism PV targets, Coma, and A~2319.

\subsubsection{The Abell 2256 cluster}
\label{2-target}

From targets in \cite{2014arXiv1412.1176K}
and recent literature
to study 
gas dynamics in a different environment, 
we select
an X-ray bright cluster A~2256 (z=0.056).
The past observations
revealed 
that this system has recently experienced a strong merger shock and is at a more violent merger stage than Coma.
This is the most suitable for the \xrism spectral mapping of gas motions and related physics,
as explained below.
{\bf
  Double peaked structures in X-ray image and galaxy dynamics:}
At the center, 
there are two substructures in the X-ray surface brightness
separating by
$3'.5$
in the sky.
The two are separated by $\sim 2000$ km s$^{-1}$ in radial velocity peaks of member galaxies as well. 
These angular and velocity separations
are unique in this system.
Almost all other X-ray bright mergers
show LOS velocities smaller than this
(e.g., 700 \kms in the Coma, 100 \kms in A3667).

\subsubsection{\suzaku discovery of gas bulk motion}
Using \suzaku CCD,
we found
a radial velocity difference between the main and subsystems
to be 1500 $\pm 300$ (statistical)  $\pm 300$ (systematic) km s$^{-1}$.
See Fig.3.
This is the first detection of gas bulk motion in a cluster.
Except for this, coma cluster \citep{2020A&A...633A..42S},
and \hitomi Perseus given above, 
no clear detection of the gas motion is found.

We found that the X-ray determined absolute redshifts of and hence the difference between the main and sub components are consistent with those of member galaxies in optical.
In fact, the X-ray redshift of the main component,
is the same as the galaxy redshift within the error.
These suggests that the X-ray emitting gas is moving together with galaxies 
as a substructure.
In addition to these, 
we found a hint of X-ray gas velocity shift with respect to the galaxy (G2)
around the sub-component at about 500 km s$^{-1}$. 
Note that this difference is within the combined errors from X-ray statistical (150 km s$^{-1}$), 
systematic (300km s$^{-1}$), and optical fitting (160 km s$^{-1}$).

\subsubsection{
Recent deep X-ray and radio observations}
\chandra observations
revealed detailed gas structures in and around
the main (P1 = CORE),
second (P2 = NW-SUB), and fainter (P3) peaks.
\cite{2020MNRAS.497.4704G}
analyzed the X-ray temperature, metallicity, and density map, 
showing clear evidence of a series of merging activities.
They examined these X-ray and radio features
and discussed a scenario
including an infalling gas from NW to the central primary.
We aim to probe LOS motions associated with this collision.
Radio observations revealed
an NW relic (the 2nd brightest one among all known systems) and a faint halo.
Several head tail galaxies (B, C, I)
also 
indicates the dynamic interaction between gas and galaxies.
We are fascinated by galaxy C,  
which has a tail extending $>500$ kpc,
but not completely aligned with the merger direction.
\cite{2003AJ....125.2393M} 
suggested that
since the tail emission is not disrupted
the region would have low turbulence.
To test this, we measure the X-ray line width (\S~\ref{2b-obj:goals}).

\section{Science Goals and Observation Plan}
\subsection{Resolve Measurement goals}
\label{2b-obj:goals}

\subsubsection{
    line shifts }
We measure a map of X-ray line redshift
and search for gas \vbulk \, 
structures associated with cluster formation.
By proposed \xrism observations,
we aim to improve and advance previous CCD results given above
by not only reducing velocity errors (\S~\ref{stat-limit})
but also
resolving velocity structure spatially.

Does the gas move together with galaxies in a merging cluster?
We compare the obtained X-ray velocities with those of galaxies
and examine if the gas components
around the merging core 
are moving together with galaxies.

We use the gas \vbulk\, map around central galaxies
to directly constrain
critical merging parameters, 
including the direction of merger axis 
and the angular moment.
There are radial velocity differences
among bright galaxies, G1, G2, and G3,
indicating the merger axis to be not in the sky plane.
Radio relic shape and polarization fraction
also indicate a tilted direction
(van Weeren et al. 2012).
This cluster has
the largest shift in optical radial velocities (\S~\ref{2-target}).
Therefore we expect to discover 
the largest angular moment in the cluster plasma.

\subsubsection{
    Line profiles and turbulent motions}
We resolve line profiles 
to measure the velocity dispersion (\vdis), 
which is 
essentially the 1D LOS component of the characteristic velocity of isotropic turbulence.

Along with gas bulk, turbulent motion
is assumed to play an important role in the galactic and larger scales (\S~\ref{1-intro:1}).
Nevertheless, 
turbulence has been more challenging to measure and hence to model theoretically.
\hitomi Perseus spectroscopy revealed
a velocity dispersion 
(\vdis)
peaked toward the central AGN at 200 km s$^{-1}$
and being uniform at 100 km s$^{-1}$ around the core.
Given the gas super-sonic motion found with \suzaku 
and other active gas features given above, 
do we expect a large \vdis\, around merging systems?
To answer this, 
we search for gas bulk and turbulent motions simultaneously
across the violent collision site.
We compare A~2256 motions with those in other places such as completely relaxed ones, 
and systematically characterize these motions.

\subsection{Pointing plan}
We aim to cover
a central $6'$ (400 kpc) region.
We estimated spectral parameters
from the \suzaku Fe-line map 
and
\chandra spectral fit
\citep{2002ApJ...565..867S}.
We expect 
the NW-SUB region to provide Fe line flux 
comparable to the core one.

\subsection{EXTEND analysis}
We use the Extend data
to map thermal structure in a large area and wide energy range.
We search for 
a hotter ($>10$~keV) thermal
and non-thermal emission components.
These spectra 
are used to constrain the inverse-Compton emission
and hence limit the magnetic field in a cluster scale
along with radio non-thermal (radio relic) imaging.
\cite{2009PASJ...61..339N}
demonstrated these analysis in a merging cluster A~3667.

\subsection{
Visibility and  technical feasibility 
}

\subsubsection{Visibility and roll angle}
The target is visible over the GO phase with a large freedom of roll angle.
We do not constrain the roll angle.
For a given observation date, we will align the Resolve angle to the cluster merger axis to cover the largest scale as much as possible.

\subsubsection{Statistical limit and exposure time request}
\label{stat-limit}
The statistical accuracy of the line shift and broadening depends on intrinsic cluster line width (\vdis).
Ota et al. (2018) showed that
$>200$ Fe line counts 
are required to achieve an accuracy better than 20\% in the gas \vdis, 
when \vdis $< 200$ \kms.
We checked these errors using the \xrism response files.
Based on these and the spectral parameters of the proposed regions from the \suzaku spectra, 
we simulated \xrism observations,
examined feasibility, and planned the observations.

\begin{table}[h]
\begin{center}
  \caption{Expected accuracy of X-ray gas velocities ($1\sigma$ statistical)
    from \cite{2018PASJ...70...51O}.
    We assume thermal 5--8~keV CIE emission.}
  \label{tbl:point}
\begin{tabular}{llll}
\hline
Fe Line & \vdis & $\Delta$\vdis  & $\Delta$\vbulk \\
(counts/100ks) & (km/s) & (km/s)  \\
\hline
200 & 0 & 30  & 30\\
200 & 200 & 30  & 75\\
50 & 0 & 50 &  30\\
50 & 200 & 50 &  100\\
\hline
\end{tabular}
\end{center}
\end{table}

\subsection{Systematic limits
}

\subsubsection{
Instrumental and astronomical errors}
In \cite{2018PASJ...70....9H}
we evaluated instrumental calibration effects 
on the velocity measurements.
We expect the Resolve
energy scale accuracy
to be reduced down to
1~eV (goal, $\sim 50$ \kms).
These levels of systematic
can be comparable when the cluster \vdis\, 
is significantly larger than the instrumental energy resolution
($\sigma \sim 2$~eV).
Note that
energy scale variations
among observations (integrated over the FOV)
and pixels
can be important.
Note also that
(1) absolute energy is critical to compare with other instruments, and
(2) relative one is to measure astronomical variations.

Based on the \hitomi analysis,
we expect other errors to be considered.
Those include
line spread function ($<0.5$~eV), 
spectral and atomic modeling errors ($<1$~eV).

\subsubsection{Spatial-spectral mixing}
Within the proposed region, 
we aim to measure
\vbulk \, 
and
\vdis \, structures
in $0'.5-1'.5$ 
and $3'$ (Resolve FOV)
spacial resolutions,
respectively, 
considering 
the pixel size of $0'.5$
and
angular resolution of $\sim 1'.3$ HPD.
Because of a relatively flat brightness distribution in this cluster, 
the PSF scattering
is less severe
than that of centrally-peaked cool core clusters.
We model spacial mixing using 
existing high spatial resolution X-ray data.
We will use a simultaneous fitting method that considers
PSF mixing
to reconstruct velocity maps
as we demonstrated for the \hitomi Perseus analysis.

\subsection{Dark matter and cosmology}
In clusters 
gas, galaxies, and associated interactions are dynamically
controlled by dark matter gravity.
Based on this assumption,
we have studied these structure formations.
In this proposal,
we focus on dark matter in an unrelated merging system
and compare it with those in a relaxed one.
Is the gas trapped at the bottom of the cluster potential or
at those of the central galaxy (cD) one?
There is a redshift difference
at about 400 km s$^{-1}$.
between the cluster mean and cD (A~2256 G1), 
which is within the feasibility above.

Accurate measurements of dark matter distribution
are required for precise cosmology.
Energies in gas motions 
could be the key uncertainty to calibrating total mass distribution,
as we discussed in \cite{2014arXiv1412.1176K}.
From the Perseus core, \hitomi revealed
little ($<10$\%) deviation from hydrostatic equilibrium.
We aim to test this basic assumption in a merging cluster.

\begin{figure}[h]
\begin{center}
  \includegraphics[width=0.8\textwidth]{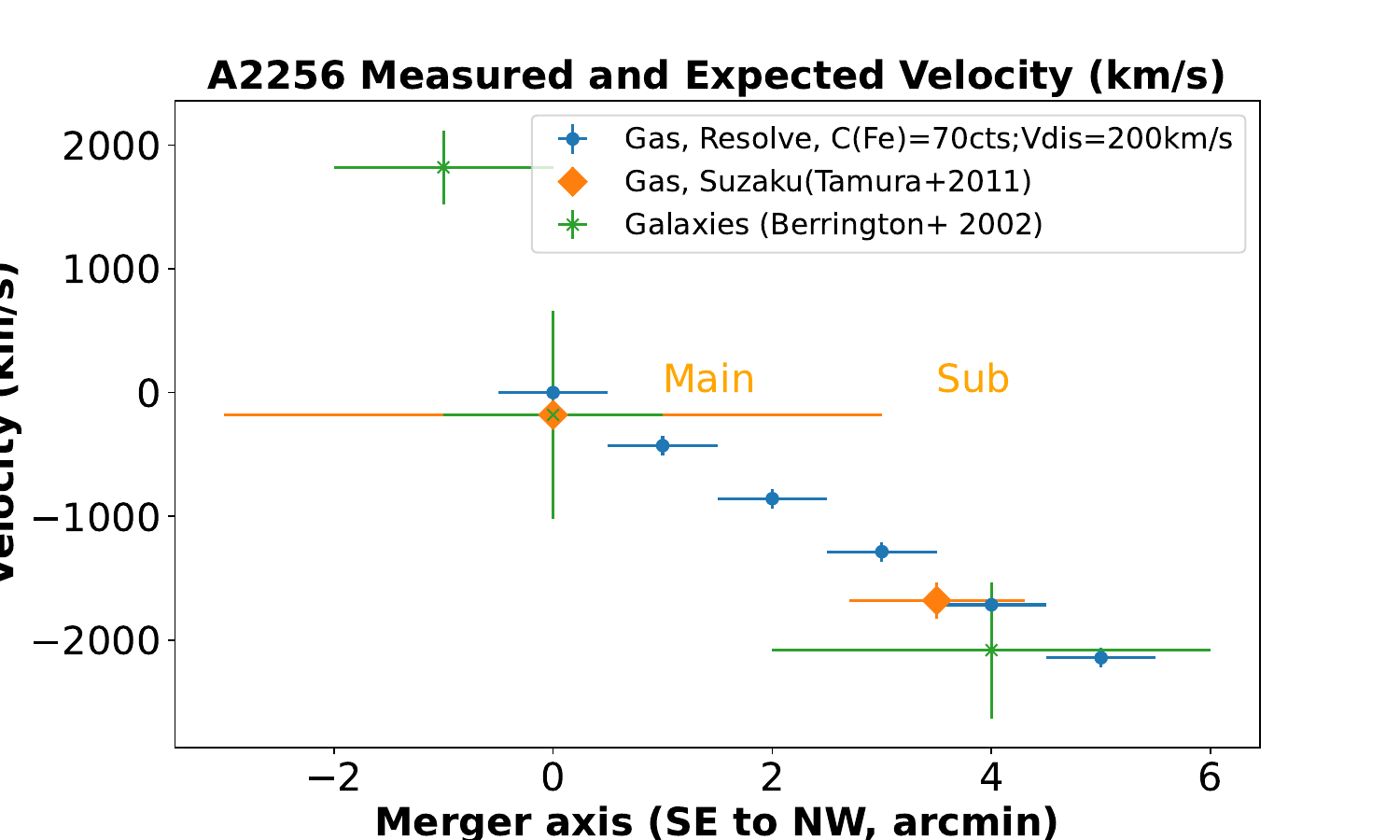}  
\caption{
  Gas and galaxies velocities from observations and XRISM/Resolve simulations.
  Resolve errors bars are statistical one ($\pm 80$\kms).
} \label{fig:velocity}
\includegraphics[width=0.8\textwidth]{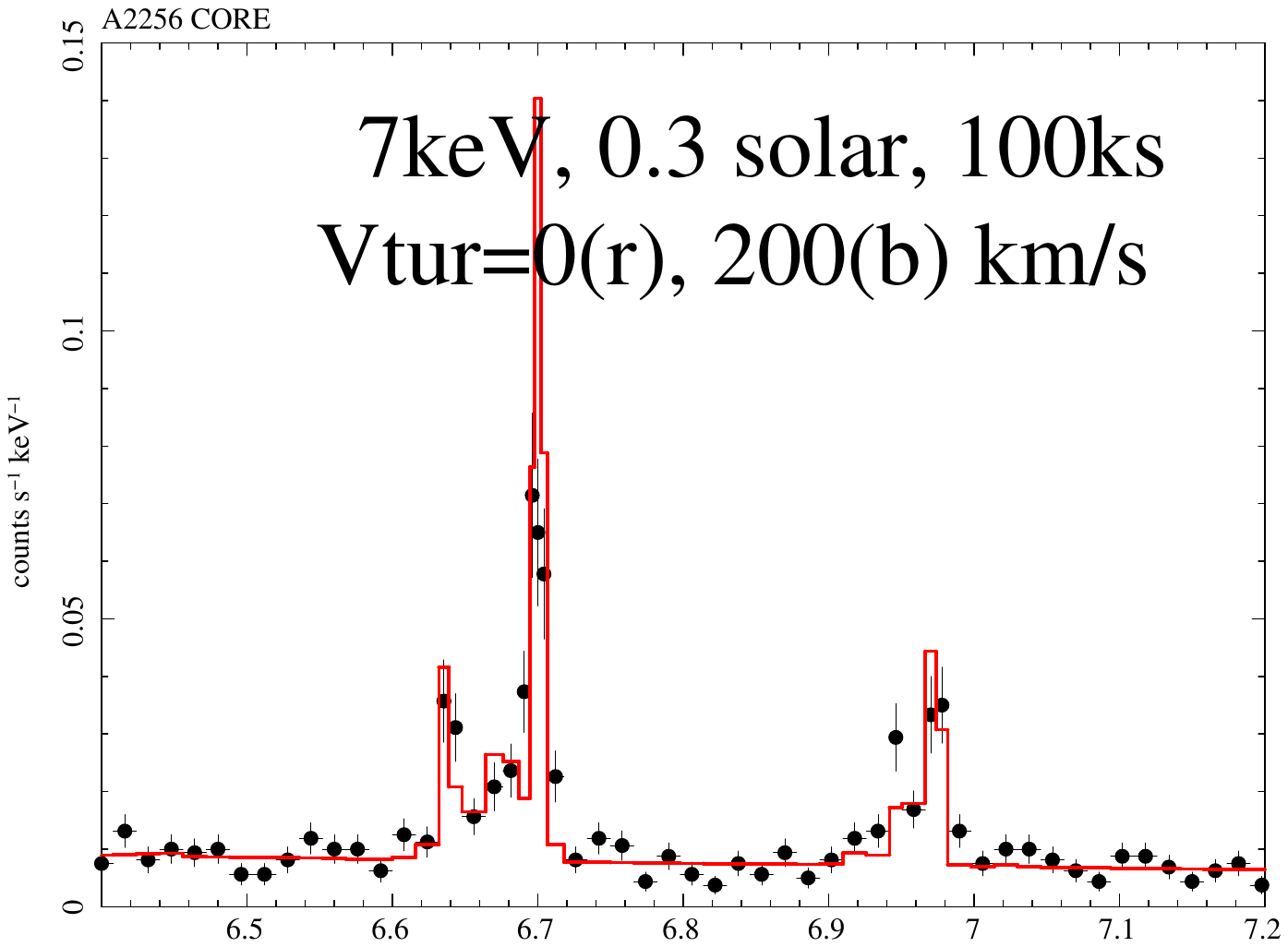}
\caption{
  Simulated \xrism spectra of
  the A2256 core component.
  Input \vdis are 0 \kms (red histogram) and 200 \kms (black points).
}
\end{center}
\label{fig:fake}
\end{figure}

\end{document}